\documentclass[10pt,twocolumn,prl,aps,amssymb,amsmath,tightenlines]{revtex4}
\usepackage{dsfont}
\usepackage{graphicx}
\usepackage{amssymb,amsfonts}

\newcommand{\re}{\mbox{$\rm e$}}

\begin{document}

\title{Outsider Trading}

\author{Dorje~C.~Brody${}^1$, Julian~Brody${}^2$,
Bernhard~K.~Meister${}^3$, and Matthew F. Parry${}^4$}

\affiliation{${}^1$Department of Mathematics, Imperial College
London, London SW7 2BZ, UK \\ ${}^2$Business Service Division,
Yahoo Japan Corporation, 9-7-1 Akasaka, Minato-ku, Tokyo, Japan \\ %107-6211 \\
${}^3$Department of Physics, Renmin University of China, Beijing,
China 100872 \\
${}^4$Department of Plant Sciences, University of Cambridge,
Cambridge CB2 3EA, UK}

%\date{\today}

\begin{abstract}
In this paper we examine inefficiencies and information disparity
in the Japanese stock market. By carefully analysing information
publicly available on the internet, an `outsider' to conventional
statistical arbitrage strategies---which are based on market 
microstructure, company releases, or analyst reports---can 
nevertheless pursue a profitable trading strategy. A large
volume of blog data is used to demonstrate the existence of an 
inefficiency in the market. An
information-based model that replicates the trading strategy is
developed to estimate the degree of information disparity.  
\end{abstract}

%\pacs{no pacs}

\maketitle

%\newpage

%\section{Introduction}
%\label{sec:1}

\noindent \textbf{1. Introduction}.
Since the dawn of history, information has always been generated
\textit{locally}; it then spreads globally by various means, often 
being lost and sometimes being
rediscovered. Nothing has fundamentally changed with the
advent of the internet. Here again, information is generated
locally on individual web sites and then, due to the potency
of content and presentation, as well as the vagaries of place
and timing, disappears into some data repository, or is picked
up and amplified, creating avalanche effects.
Nowadays information is often posted initially on blogs and
twitter accounts, or discussed on bulletin boards, and only
subsequently, with some delay, reaches the traditional media
as represented by newspapers and television. This 
dissemination from a small
to a wider circle of viewers is also of interest in the
financial market context, because as knowledge spreads, it starts
influencing investment decisions. We demonstrate in this
paper that by capturing these trends at an early
stage of information diffusion in a systematic and quantitative
manner it is possible to construct a superior trading strategy, 
thus establishing the existence of market inefficiencies.

A closely related issue to information extraction in
financial markets is the valuation of information.
Suppose one is in possession of a piece of information,
deemed valuable, that one wishes to monetise. How does
one \textit{price} information, when information is viewed as
a tradable asset?  For instance, consider the
information that the price of a given stock will move up the
following day with 75\% likelihood. Leaving aside issues to
do with insider trading for the moment, if one was to `sell'
this piece of information, how should one set a fair price?
Evidently, in this example the price depends on a number
of market factors, such as market impact. It also depends
crucially on whether this information provision is a one-off
event or whether such information will be supplied on a
regular basis. All these issues make it virtually impossible
to arrive at the notion of a `fair price' of information. It is
nevertheless possible to associate a \textit{rate of
return} with the use of information, as we shall
show here.

In the efficient market theory `all' the publicly available
information is incorporated in the price by the marginal
investor. This bold statement, which comes in various
forms, has often been criticised in the literature (e.g., 
Grossman and Stiglitz 1980). Often there is an abundance 
of valuable information that is widely accessible to the 
whole market but from which not everyone has the 
resources or analytic capability to extract useful signals. 
Indeed, not even the so-called `marginal' investor appears 
to exploit this additional data. The important point is that 
the distribution of information is never homogeneous 
because the ability to extract something useful is 
inhomogeneous across different market agents.

To establish a relationship between information and
investment return, we must first identify what is meant by
information. In financial markets information consists of two 
parts: signal and noise. By `signal' we mean components of 
information that are dependent on the actual return of, say, 
an investment; whereas by `noise' we mean components of 
information that are statistically independent of the actual 
return of that investment. Both components have direct impact on 
price dynamics, but it is ultimately the signal component that 
determines the realised value of the return. Thus, given this noisy 
information, market participants try their best to estimate the signal; 
this estimate (in a suitably defined sense discussed below) in turn
determines the random dynamics of the associated price process.

In many cases signal and noise are superimposed in an additive
fashion. In other words, there are essentially two unknowns,
`signal' and `noise', and one known, `signal plus noise'. The rate 
at which the signal is revealed to the market then determines the
signal-to-noise ratio. The kind of information inhomogeneity
discussed above therefore arises primarily from the fact that 
different agents have different signal-to-noise ratios. With further
refinements, however, one finds that signal-to-noise ratio is itself
rarely known in financial markets, i.e. it is what one might call
a \textit{known unknown}. Yet, it is the signal-to-noise ratio
that directly affects the performance of an investment. Hence we
can determine the relative ratios of signal-to-noise ratios of
different agents from their performances. This is one objective of
the present paper. We examine the ratio of
two signal-to-noise ratios; one for the market as a whole, and
one for an internet-search based strategy.

Our choice for using an internet-search based strategy, as 
a comparison against the market, should be evident: most 
information circulates via the internet. Unlike traditional 
investment firms, large internet search engines, by their 
very nature and in spite of being `outsiders' to financial 
markets, are well positioned to extract signals from large 
data sets. From the viewpoint of internet search engines, the 
kind of analysis discussed here also has a profound implication. 
One of the key difficulties in the business of information provision 
is in the quantitative assessment of the validity and quality of the 
search engines or other recommendation tools. However, we now 
recognise that financial market dynamics provide a suitable 
testing ground, and one with rapid feedback. For example, a 
``celebrity popularity engine'' offered by internet companies, 
useful to advertisers, can be applied to individual companies; 
the quality of the engine, which otherwise would have been 
difficult to assess, can now be tested instantly against the future 
movements of the corresponding stock prices.

We have therefore taken a large number of blog articles from the internet, 
applied natural language processing (NLP) to convert numerous 
texts into numerical sentiment indices for individual listed companies, 
and then developed a trading strategy that converts the sentiment 
indices into portfolio positions. The results show the existence of an
astonishing inefficiency in a highly liquid equity market. We also 
construct a theoretical model, within the information-based
asset pricing framework of Brody-Hughston-Macrina (BHM), for the
characterisation of the strategy. The model has the advantage that
the ratio of the signal-to-noise ratios between the informed outsider
and the general market can be estimated from the investment
performance.

\vspace{0.15cm}

\noindent \textbf{2. Information and asset price}. To understand the
interplay between information and asset price, we must first step
back from the conventional approach in quantitative finance, and begin
by identifying the main sources for price movements at a
phenomenological level.  After a little reflection it should not be difficult
to identify two important factors, namely, risk preference and available
information. To understand these two factors we list two different
scenarios: (i) I would have bought the new Toyota car, had I not lost
my job; (ii) I would have bought the new Toyota car, had I not read
the news of the recall. In case (i) the assessment of the worthiness of
the product has not changed, but the purchase decision has
nevertheless been affected by the changes in one's appetite toward
risk; whereas in case (ii) the assessment of the worthiness of the
product has changed due to the arrival of new information.

It is often argued that the price dynamics is generated by supply and
demand; this is indeed so, but it has to be noted that a large part of
supply and demand in financial markets is induced by the arrival of
information (for example, an announcement of a substantial profit
leading to high demand for company shares). We thus take the
view that the traditional `supply and demand' argument is in fact mostly
the symptom and not the cause, at least in the case of highly liquid
financial instruments. 

As regards changes in risk preference, at the individual level this can
be relatively volatile, but averaged over the market the volatility will be 
reduced. On the other hand, the flow of information is significantly more
dynamic and volatile. It is common for a dynamical system to depend
on fast moving and slowly moving variables; in the case of a financial
market, information is the fast moving and risk preference is the slowly
moving variable. For our strategy, the changes in overall 
risk preference have little impact, because we only test market neutral 
strategies that have no exposure to the overall risk preference of the 
market. Therefore, our first
simplifying assumption is to regard market risk preference as fixed,
and focus attention on the structure of information. Phrased in more
technical terms, we will assume that the pricing measure is given
once and for all, and we shall construct the market filtration from
the outset, which will be used to derive the price process. This is
in line with the BHM approach introduced in Brody \textit{et al}. 
(2007, 2008), which will now be reviewed briefly.

Consider an elementary asset that pays a single dividend $X$ at
time $T$ (e.g., a credit-risky discount bond). We assume that there
is an established pricing measure ${\mathbb Q}$, under which the
random cash flow $X$ has the \textit{a priori} density $p(x)$. In this
case, market participants are concerned about the realised value
of $X$. In particular, the risk-adjusted view of the 
market today
about the cash flow is represented by the \textit{a priori} density
$p(x)$. By tomorrow, however, the market will obtain additional noisy
information, based on which the market will update its view,
represented in the form of an \textit{a posteriori} density for $X$.
This information consists of two components;
signal and noise. Although the signal-to-noise
ratio is generally unknown, and furthermore it will change in time, 
let us assume for simplicity that it is
known to the market, and that it is given by a constant $\sigma$.
We also assume for the moment that the market is efficient in the
sense that all available information is used in the determination
of the price today. Hence there is no residual noise today.
Likewise, the noise will vanish at time $T$ when the value of $X$ 
is revealed for sure. To keep the matter simple, we model the
noise term by the simplest Gaussian process that vanishes at
time $0$ and time $T$---the Brownian bridge process
$\{\beta_{tT}\}$ over the time interval $[0,T]$. Therefore, our
choice for the information is 
\begin{eqnarray}
\xi_t = \sigma X t + \beta_{tT}. \label{eq:1}
\end{eqnarray}
The market filtration $\{{\mathcal F}_t\}$ is thus generated by the
knowledge $\{\xi_s\}_{0\leq s\leq t}$ of the information process.

If we write $P_{tT}$ for the discount function, and assume that
it is deterministic, then the price at time $t$ of the asset is
determined by
$S_t = P_{tT} {\mathbb E}[X|{\mathcal F}_t]$. 
A short calculation then shows that the price process is given by
\begin{eqnarray}
S_t = P_{tT} \frac{\int_0^\infty x p(x) \re^{\frac{T}{T-t}\left(\sigma
x\xi_t-\frac{1}{2}\sigma^2 x^2 t\right)}{\rm d}x}{\int_0^\infty p(x)
\re^{\frac{T}{T-t}\left(\sigma
x\xi_t-\frac{1}{2}\sigma^2 x^2 t\right)}{\rm d}x} .  \label{eq:3}
\end{eqnarray}
We see therefore that in the BHM framework it is possible to
\textit{derive} the price process in a manner that replicates
how price processes are generated in the first place via flow
of information. In spite of the various simplifying assumptions,
the resulting price process (\ref{eq:3}) is very rich and
possesses many desirable features. Perhaps the most notable
from a practical point of view is the fact that the pricing and the
hedging of elementary contingent claims are made easy.

\vspace{0.15cm}

\noindent \textbf{3. Modelling the informed outsider}. Within the
BHM framework it is straightforward to model the information
disparity seen in the market. Indeed, it has been shown in
Brody \textit{et al}. (2009) that if there is an informed trader in
the market who has access not only to the market information
(\ref{eq:1}) but also to an additional information source $\xi'_t
=\sigma' Xt+\beta'_{tT}$, then the informed trader can exploit
the information to generate statistical arbitrage. Here we
shall modify the setup considered therein 
so as to replicate the trading strategy that we have developed
by use of data taken from the internet, and calibrate some of
the model parameters. In this manner we are able to test the 
performance of internet-based recommendation or rating 
engines from investment performances.

Our modelling setup can be summarised as follows. We let $X$
be a binary random variable taking the values $\{0,1\}$, where
$1$ represents price moving up by a unit over the period $[0,T]$ 
and $0$ represents price moving down by a unit over the same 
period. At time $0$ both the market and the informed trader share 
the same information about the value of $X$, represented by the
\textit{a priori} probabilities $(p,1-p)$. The informed trader,
however, begins to gather information from the internet,
using text and data mining; whereas the general market gathers 
information through more widely accessible sources such as 
newspaper articles and financial reports. We let $\xi_t$ of
(\ref{eq:1}) represent the market information process, and
$\xi'_t =\sigma' Xt+\beta'_{tT}$ represent the extra information
gathered from the internet, where the two noises $\{\beta_{tT}\}$
and $\{\beta'_{tT}\}$ may be dependent, with correlation $\rho$.
It is shown in Brody \textit{et al}. (2009) that in the case of multiple
information sources the knowledge of the informed trader can
be represented in the form of a single effective information
process
\begin{eqnarray}
{\hat\xi}_t = {\hat\sigma} X t + {\hat\beta}_{tT}, \label{eq:4}
\end{eqnarray}
where ${\hat\sigma}^2 = (\sigma^2-2\rho\sigma\sigma'+
\sigma^{\prime 2})/(1-\rho^2)$, and
\begin{eqnarray}
{\hat\beta}_{tT} =
\frac{\sigma-\rho\sigma'}{{\hat\sigma}(1-\rho^2)} \beta_{tT}  +
\frac{\sigma'-\rho\sigma}{{\hat\sigma}(1-\rho^2)}
\beta'_{tT} . \label{eq:5}
\end{eqnarray}
Therefore, the effective signal-to-noise ratio for the informed
trader is given by ${\hat\sigma}$, which can be compared
against the market signal-to-noise ratio $\sigma$.

At time $T/2$ both the market and the informed trader have
accumulated noisy information, based on which they evaluate
the \textit{a posteriori} probabilities, $p_m$ and $p_i$,
respectively, that $X=1$. The trading strategy is as follows.
If the \textit{a posteriori} probability is larger than the threshold
value $K_+$ then take a long position by the amount ${\hat X}_t$;
if the \textit{a posteriori} probability is smaller than the threshold
value $K_-$ then take a short position by the amount ${\hat X}_t$,
where ${\hat X}_t={\mathbb E}_t[X]$ is the expectation of $X$
using the market filtration. The position is then held till time $T$,
at which point the profit or loss is made because the value of $X$
is now revealed. Also at time $T$ the next observation for the value
of the random variable representing whether the asset
price moves up or down over the interval $[T,2T]$ begins, and
the same strategy is repeated over and over. Our model thus
makes an implicit simplifying assumption that the magnitude of
the stock volatility over the range $[nT,(n+1)T]$ is independent
of the value of $n$.

Both the market and the informed trader employ the same
strategy, but the informed trader \textit{on average} makes
better estimates for the realised value of $X$, thus statistically
obtaining a higher rate of return than the market. The risk-neutral
valuation of the market position can be made straightforwardly,
because the resulting cash flow is given by $(X-{\hat X}_t)
({\mathds 1}\{{\hat X}_t\!>\!K_+\}-{\mathds 1}\{{\hat X}_t\!<\!K_-\})$.
By a change of measure technique introduced in Brody
\textit{et al}. (2007) one can show that the value of the strategy
is given by a formula analogous to the Black-Scholes option
pricing formula. The valuation of the position of the informed
trader is less obvious, although one can show
that the expected P\&L difference is positive, leading to
a statistical arbitrage opportunity.

\vspace{0.15cm}

\noindent \textbf{4. Implementation and calibration}.
We have implemented the strategy using publicly available 
information sources. Specifically, we have gathered the 
totality of Japanese blog articles since 2006 and used them 
as our sole information source. In 2009, nearly 20 million
Japanese blog articles appeared on the internet, making a
daily average of around 50,000 articles. Each blog article is
weighted by its relevance (e.g., page views). Those with
insufficient weight are regarded as `pure noise' and have
been discarded from the analysis.
%, thus leaving on average 6,858
%articles per day (for 2009) for the analysis.

Natural language processing (NLP) technology of Yahoo
Japan Corporation and Yahoo Japan Research Institute has
been applied to analyse company specific comments of the
listed companies. The NLP classifies whether the comments
are positive, neutral, or negative; this classification is then
used to establish sentiment index for each company. Based
on the sentiment index, a trading strategy, analogous to the
one described above, is developed. The idea can be 
illustrated as follows. If many people write complimentary
remarks about a new product released by a given company 
then it is likely that sales of the product will go up, leading
to an increase in its share price.

%%%%%%%%%%%%%%%%%%%%%%%%%%%
\begin{figure}[th]
\begin{center}%\vspace{-2.5cm}
  \includegraphics[scale=0.50]{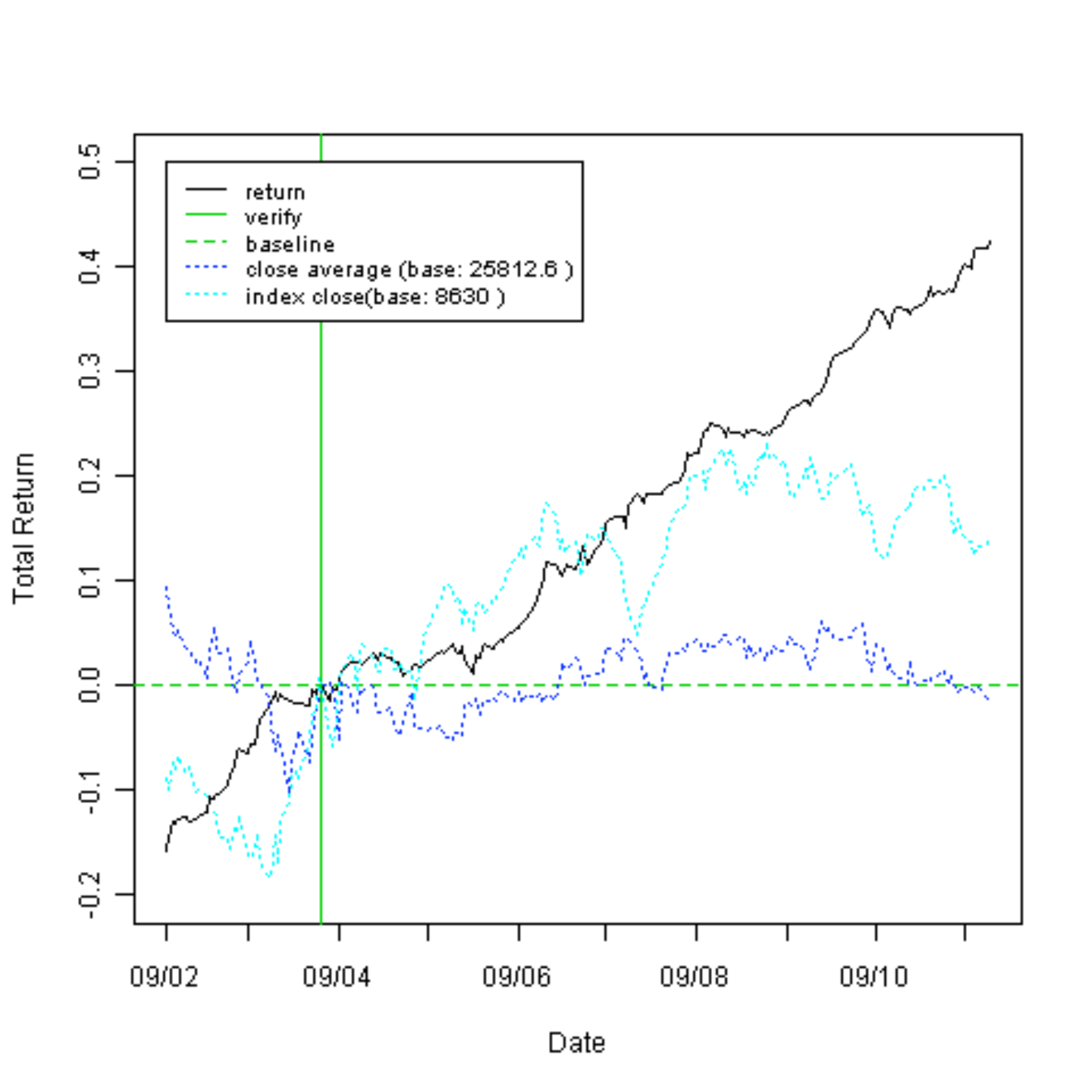}
  %\vspace{-2.5cm}
  \caption{\textit{Information-based trading}. The blog sentiment
  data is used to create a trading strategy for the relevant stocks. 
  The performance (total return) that results from the strategy is
  shown in the solid black line. The dark blue dashed
  line represents the average stock prices; the light
  blue dashed line the Nikkei 225 Index. The `learning' period
  for optimisation corresponds to the left of the vertical green line;
  the strategy is applied over the seven month period starting in
  the late March 2009. Our information-based strategy yields
  over 40\% return for the seven month period.
  \label{fig:1}
  }
\end{center}
\end{figure}
%%%%%%%%%%%%%%%%%%%%%%%%%%%%

The strategy has been optimised using the data from 2008 to
early 2009 (for example, the choice of the threshold values
$K_\pm$), and applied for the seven month period from April
2009. Specifically, for the analysis presented here we have
considered 10 companies for whom the average numbers of
blog comments are highest. In order to obtain a conservative
estimate for the ratio ${\hat\sigma}/\sigma$, and also to reduce 
exposure to the market risk preference, we have adopted
a long-short strategy against the Nikkei 225 Index. The result
of the strategy, as well as the average stock prices of the active
names and the Nikkei 225 Index, are shown in figure~\ref{fig:1}.

%%%%%%%%%%%%%%%%%%%%%%%%%%%
\begin{figure}[th]
\begin{center}%\vspace{-2.5cm}
  \includegraphics[scale=0.40]{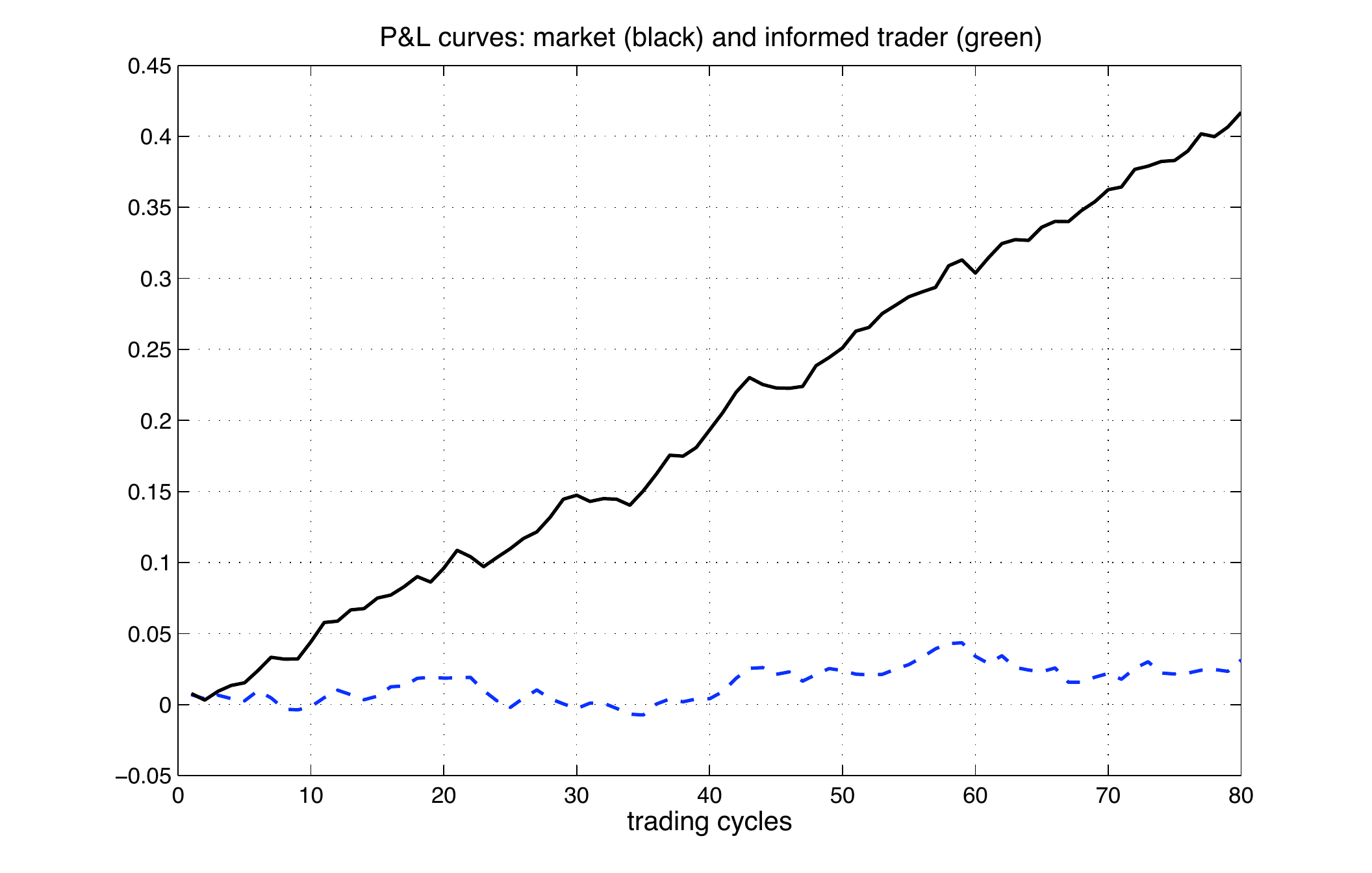}
  %\vspace{-2.5cm}
  \caption{\textit{Simulation of the strategy}. We have
  run the strategy for the informed trader (black solid line)
  and that for the market (blue dashed line), and taken the
  average over 5,000 sample paths. Parameter values are
  set as $\rho=0.1$, $\sigma=0.2$, $\sigma'=0.48$, and
  hence ${\hat\sigma} = 0.50$.
  \label{fig:2}
  }
\end{center}
\end{figure}
%%%%%%%%%%%%%%%%%%%%%%%%%%%%

To estimate the ratio ${\hat\sigma}/\sigma$ we have
simulated the strategy numerically. Because we do not
yet have a suitable method of estimating the correlation
$\rho$ between the noise in the blog sentiments and the
noise for market investors, we can only give a range for
this estimate. Fortunately, however, we found that the
range is relatively narrow:
\begin{eqnarray}
2.4 \lesssim \frac{{\hat\sigma}}{\sigma} \lesssim 2.6.
\end{eqnarray}
The simulation results associated with the choice $\rho=0.1$
are shown in figure~\ref{fig:2}.

\vspace{0.15cm}

\noindent \textbf{5. Discussion}. We have successfully extracted
a trading signal from the abundant data accessible on the internet. 
By applying the results to the stock market, we were able to assess 
the performance of the information extraction and provision engine. 
The results have identified perhaps a surprising level of apparent 
inefficiency even in a highly liquid equity market, indicating the 
degree of information inhomogeneity.

It is of course well documented that asset prices in financial 
markets respond to the unravelling of information (e.g., Engle 
and Ng 1993; Andersen \textit{et al}. 2007). Indeed, the 
realisation that information filtering and communication is the 
key for grasping social sciences such as economics has been 
recognised since Wiener (1954). Our analysis differs sharply 
from previous work carried out in this area in that we explicitly 
identify the existence of information disparity and derive an estimate for
how much more the rate of information extraction could have been
enhanced had the market been truly efficient. In contrast with Google  
Finance, for instance, that provides a postmortem analysis of
the relation between large price moves and revelations of news
items, our informed trader is able to exploit additional information  
sources to anticipate price moves. 

The analysis reported in this paper is naturally of interest to
statistical arbitrage funds, because the strategy is orthogonal 
to conventional strategies that rely on, for example, microstructure.
On the other hand, from the viewpoint of an internet search engine,
one might envisage a scenario whereby individual investors
purchasing `signal' from information providers and making their own
investments. Such a model, however, is unlikely to be sustainable,
because if the signal is circulated broadly, it ceases to 
remain useful. As Wiener emphasises, concentration of useful
information is intrinsically unstable due to the second law (Wiener
1954). The only way in which information can be spontaneously 
concentrated, at least momentarily, is via innovation. It is interesting 
therefore to reflect on the fact that in
spite of the enhancement of technology in improving the method
of information gathering and provision, whose purpose \textit{a
priori} goes against the second law, ultimately such
developments can only result in enforcing the compliance with
the second law. As a result, in the
long run the second law will enhance the `efficiency' of 
financial markets, but maybe also, paradoxically, the instability
of financial markets, because in a noise-dominated market, the 
revelation of the true signal has a significant impact.

\vspace{0.15cm}

\begin{acknowledgments}
The authors thank Robyn Friedman for stimulating discussion.
The opinions expressed in this article are those of the authors. 
Email: dorje@imperial.ac.uk${}^*$, brodyj@yahoo-corp.jp, 
b.meister@imperial.ac.uk${}^*$, and m.parry@statslab.cam.ac.uk 
(${}^*$ corresponding authors).
\end{acknowledgments}

\end{document}